\documentstyle[12pt,epsf,rotate]{article}

\newcommand{\be}{\begin{equation}}
\newcommand{\ee}{\end{equation}}
\newcommand{\ba}{\begin{eqnarray}}
\newcommand{\ea}{\end{eqnarray}}
\newcommand{\NL}{\nonumber \\ & & \nonumber \\}

\newcommand{\LO}{\tau}
\newcommand{\grad}[1]{\frac{\partial}{\partial #1}}
\newcommand{\ddL}{ \left( \grad{\LO} + v \grad{r} \right) }

\newcommand{\sh}{{\rm sinh}\,}

\newcommand{\vs}{\vspace{0.5cm}}

\begin{document}


\baselineskip 14.5pt
\parindent=0cm
\parskip 3mm



\begin{center}
{\bf \huge 
 Generating new solutions for relativistic transverse flow
 at the softest point
}

\vspace{1.0cm}

{\sc Tam\'as S. Bir\'o\footnote{e-mail: tsbiro@sunserv.kfki.hu
 ; \quad 
 http://sgi30.rmki.kfki.hu/$\tilde{~}$tsbiro/ 
}  
}  

\end{center}

\centerline{\small Reseearch Institute for Particle and Nuclear Physics}
\centerline{\small             H-1525 Budapest, P.O.Box 49, HUngary}

\vspace{0.5cm}


\vspace{1.0cm}
\begin{small}
{\bf Abstract:} 
Using the method of prolongation we generate new stationary solutions
from a recently obtained simple particular solution for
relativistic transverse flow with cylindrical symmetry in 1+3 dimension.
This is an extension of the 
longitudinal Bjorken flow ansatz and can be applied to situations
during a first order phase transition in a fast expanding 
system. 
The prolongated solution allows us to trace back the flow profile
from any transverse flow conjectured at the end of the
phase transition.

\end{small}

\vspace{1.0cm}
\noindent
{\bf \sc Introduction}
\vspace{0.5cm}

Hydrodynamics often allow for nonrelativistic scaling solutions.
Relativistic flow, however, seemed long  to be an exception: besides
Bjorken's 1+1 dimensional ansatz and the spherically symmetric
relativistic expansion, no analytical solution was known \cite{Csernai}.

In a recent paper \cite{Biro}
 we presented an extension of Bjorken's ansatz \cite{Bjorken}
for longitudinally and transversally relativistic
flow patterns with cylindrical symmetry in 1+3 dimensions.
It satisfies $u^{\mu}\partial_{\mu} u^{\nu}=0$ as well as
the original Bjorken ansatz.
This analytical solution of the flow equations of a perfect fluid
is in particular valid for physical situations when the sound velocity is zero,
$c_s^2 = dp/d\epsilon = 0$,
with energy density $\epsilon$ and pressure $p(\epsilon)$.

In particular this happens during a first order phase transition, 
the pressure is constant while the energy density changes
(in heavy ion collisons increases and drops again). This
should, in principle,  be signalled by a vanishing sound velocity.
A remnant of this effect in finite size, finite time
transitions might be a softest point of the equation of
state, where $c_s^2$ is minimal. In fact, this has been suggested
as a signal of phase transition by Shuryak\cite{Shuryak}, and investigated
numerically in several recent works \cite{Rischke, Dumitru}.

In this paper we generalize that simple analytic solution further
by exploring the symmetries of the nonlinear partial
differential equation determining the flow at the
softest point (the relativistic Euler's equation).
The method of prolongation ensures us that all possible
transformations of dependent and independent variables can be
found which comply with the equations to be solved by
solving linear partial differential equations only.
Then from any simple (or trivial) particular solution
of the original nonlinear problem several new classes of
solutions can be generated. In this article we present
the simple solution published recently, we review the
method of prolongation and obtain a general class of transverse flow
patterns for the relativistic problem at the softest point.
We use this solution for calculating backwards from a conjectured
final state at CERN SPS \cite{Ster}.

Nonrelativistic analytic solution has been also given several times,
with respect to heavy ions see \cite{Zim,Cs1,Cs2}.

\vspace{1.0cm}
\noindent
{\bf \sc Simple relativistic transverse flow at the softest point}
\vspace{0.5cm}

The 1+1 dimensional Bjorken flow four-velocity is a normalized,
timelike vector. It is natural to choose this as the first of
our comoving frame basis vectors (vierbein). 
The three further, spacelike vectors
will be constructed orthogonal to this, separating the two
transverse directions. This basis fits excellently to a
cylindrical symmetry and to longitudinally extreme relativistic
flow.
\ba
e_{\mu}^0 = \left( {t \over \LO}, {z \over \LO}, 0, 0 \right), & \qquad & 
e_{\mu}^1 = \left( {z \over \LO}, {t \over \LO}, 0, 0 \right), \NL
e_{\mu}^2 = \left( 0, 0, {x \over r}, {y \over r},  \right), & \qquad &
e_{\mu}^3 = \left( 0, 0,-{y \over r}, {x \over r},  \right), 
\label{C_BASIS}
\ea

with $t$ time coordinate and $z$ longitudinal (beam-along)
coordinate, $x$ and $y$ transverse, cartesian coordinates.
The cylindrical radius is given by,
$r = \sqrt{ x^2 + y^2}$,
and $\LO$ is the {\em longitudinal proper time}:
$\LO = \sqrt{ t^2 - z^2}$.

\vs
We consider ideal fluids (``dry water''), where the energy momentum tensor
is given by
\be
T_{\mu\nu} = (\epsilon + p) u_{\mu} u_{\nu} - p g_{\mu\nu},
\label{T_MU_NU}
\ee
and the equation of state is given in the form of $p(\epsilon)$.
The ansatz for an almost boost invariant flow with some transverse,
cylindrically symmetric component is then given by
\be
u_{\mu} = \gamma \left( e_{\mu}^0 + v e_{\mu}^2 \right),
\label{U_ANSATZ}
\ee
using the Lorentz factor $\gamma = (1-v^2)^{-1/2}$.
This four-velocity is normalized to one:
\be
u_{\mu} u^{\mu} = \gamma^2 - \gamma^2 v^2 = 1.
\ee
In a recent publication we formulated Euler's equation
in terms of this ansatz and coordinates. For a situation with
zero comoving gradient of the pressure it has been simplified to a sole
partial differential equation for the transverse flow
velocity component $v(\LO,r)$:

\be
 \ddL v = 0.
\label{ANAL_FLOW}
\ee
This equation is valid even for relativistic transverse flow, it is
a quasilinear partial differential equation. 
The separable solution,$v(r,\LO) = a(r) b(\LO)$,
which is regular at the cylindrical axis $r=0$ is given by
\be
v = \frac{r}{\LO}.
\label{TRV_SOL}
\ee
In a recent paper we then explored consequences of such a linear
transverse velocity profile for the time evolution of the
energy density at constant pressure, i.e. the quark matter to hadron matter
phase conversion\cite{Biro}.

\vs
Our concern now is to construct a more general solution of
eq.(\ref{ANAL_FLOW}), which is able to fit {\em any}
flow profile at a given longitudinal time
\be
v(\LO_0,r) = v_0(r).
\label{INIT}
\ee

In order to explore the quark matter hadron matter phase
conversion we utilize simple bag model equation of state,
as was done in \cite{Biro}.
The energy density of $\chi$ part quark matter and $(1-\chi)$ hadron matter 
is then given by
\be
e = \left( \sigma_q T_c^4 + B \right) \chi + \sigma_h T_c^4 (1-\chi),
\ee
at the phase equilibrium temperature $T_c$, determined from the 
equity of quark and hadron pressure. 
The resulting evolution equation turns to be
\be
\partial_{\mu} ( \chi u^{\mu} ) + \nu \partial_{\mu}u^{\mu} = 0,
\label{CONVERT}
\ee
with
\be
\nu = \frac{4}{3} \frac{\sigma_h}{\sigma_q - \sigma_h}.
\ee
Here the ratio of Stefan-Boltzmann constants $\sigma_{q,h}$ depends
only on the relative number of the effective degrees of freedom.
By comparing a relativistic pion gas to quark gluon plasma
$\sigma_h = 3$ and $\sigma_q = 37$ and one obtains $\nu \approx 0.12.$

\vs
Now using the flow ansatz discussed above and eq.(\ref{TRV_SOL})
one arrives at
\be
\ddL \chi + (\chi+\nu) \left( \frac{\partial v}{\partial r} + 
\frac{v}{r} + \frac{1}{\LO} \right) = 0.
\ee
Furthermore, expressing the quark matter part as
\be
\chi(\LO,r) = - \nu + \frac{1}{\LO r} Z(\LO, r),
\label{Zdef}
\ee
we obtain a simpler equation for $Z$,
\be
\frac{\partial }{\partial \LO} Z + \frac{\partial }{\partial r} (v Z) = 0,
\label{Z_EQ}
\label{Zconv}
\ee
which formally resembles a one-dimensional conservation equation
with density $Z$ and convective current $vZ$.
It is noteworthy that for any solution $Z$ of this equation (\ref{Z_EQ})
$f(v) Z$ is also a solution with $f(v)$ being an arbitrary (but differentiable)
function of the transverse velocity $v$.
This freedom may be
used for fitting a quark matter percentage profile at the same
time when the transverse velocity profile is known, especially
the $\chi=0$ (end of phase conversion) situation may be traced back
up to a time when somewhere $\chi=1$ (full quark matter) is reached.


\vs
Eq.(\ref{Zconv}) is formally a conservation equation,
whose solution, $Z(\LO,r)$ satisfies
\be
Q(\LO) = \int_0^{R(\LO)} \nolimits  Z(\LO,r) \, dr =
Q(\LO_0).
\ee
The longitudinal time derivative of this expression vanishes leading to
\be
\frac{dQ}{d\LO} = \dot{R}(\LO) Z(\LO,r) + 
\int_0^{R(\LO)} \nolimits  \frac{\partial Z}{\partial \LO} \, dr = 0.
\ee
Replacing $\partial Z / \partial \LO$ from eq.(\ref{Zconv}) we arrive
at
\be
\left[ \dot{R}(\LO) - v(\LO,R(\LO))   \right] Z(\LO,R(\LO)) = 0
\ee
if we take into account the regularity condition $v(\LO,0)=0$.
This is an evolution equation for the transverse radius $R(\LO)$
of the phase mixture region where $Z \ne 0$,
\be
\dot{R}(\LO) \, = \, v(\LO,R(\LO)).
\ee
It is a first order ordinary differential equation which can be
easily solved once $v(\LO,r)$ is known.

\vs
The conserved quantity $Q(\LO)$ can be related to the spatial
average of the phase mixture ratio by using eq.(\ref{Zdef}):
\be
\langle \chi \rangle = \frac{2}{R^2} \int_0^R \, \chi(\LO,r) r dr.
\ee
Finally we obtain
\be
\langle \chi \rangle =  - \nu + \frac{2}{\LO R^2(\LO)} Q(\LO).
\ee


\vs
In the followings we turn to a brief presentation of the method
of prolongation, which shall be used for generalizing the
particular relativistic transverse flow solution presented
in \cite{Biro}.

\vspace{1.0cm}
\noindent
{\bf \sc The method of prolongation}
\vspace{0.5cm}

The method of prolongation ``prolonges'' symmetries of
a (system of) nonlinear partial differential equations,
to transformations of dependent and independent variables
and all partial derivatives of the dependent variables up to a degree
less then the rank of the equation itself. 
For an incomplete list of references
about the application of this method in physics see \cite{prolong}.

\vs
Let us generally denote the set of independent variables by $x$,
the dependent ones by $v$, partial derivatives by $v_x$, $v_{xx}$ etc.
The equation(s) to be solved let be given by
\be
\Delta(x, v, v_x, \ldots ) = 0.
\label{PD}
\ee
Transformations on $x$ and $v$ (generating corresponding
transformations on the partial derivatives) may let this equation
remain valid in the transformed variables as well. These transformations
are symmetries of the equation, and they can be viewed as integrals
of infinitesimal transformations generated by the vector field
(Killing vectors)
\be
K = \Phi(u,x) \frac{\partial}{\partial v} + \xi (v,x) 
\frac{\partial}{\partial x}.
\label{KV}
\ee


\vs
This relation can be easily understood by inspecting infinitesimal
transformations,
\ba
\tilde{v} &=& v + \varepsilon \Phi,  \NL
\tilde{x} &=& x + \varepsilon \xi.
\ea
In this case the derivative transforms to
\be
\frac{d\tilde{v}}{d\tilde{x}} = 
\frac{\frac{dv}{dx}+\varepsilon\frac{d\Phi}{dx}}{1+\varepsilon\frac{d\xi}{dx}}
= \frac{dv}{dx} + \varepsilon \left(\frac{d\Phi}{dx} 
- \frac{d\xi}{dx} \frac{dv}{dx}  \right) 
+ {\cal O}(\varepsilon^2)
\ee
The total derivatives are
\ba
\frac{d\Phi}{dx} &=& \Phi_v v_x + \Phi_x, \NL
\frac{d\xi}{dx} &=&  \xi_v v_x + \xi_x.
\ea
The differential equation, $\Delta = 0$ can be expanded around the
original solution
\be
\Delta(\tilde{v},\tilde{x},\tilde{v}_x) = \Delta(v,x,v_x) +
\varepsilon \Phi \frac{\partial}{\partial v} \Delta
+ \varepsilon \xi \frac{\partial}{\partial x} \Delta
+ \varepsilon \left( \frac{d\Phi}{dx}-\frac{d\xi}{dx} v_x \right) 
\frac{\partial}{\partial v_x} \Delta
+ {\cal O}(\varepsilon^2).
\ee
The second and third term of ${\cal O}(\varepsilon)$ in the above
expansion resembles $\hat{K}\Delta$, but all terms to this order
constitute a Killing field in the extended (prolongated) space
spanned by $v, x$ and $v_x$. This is the (first) prolongation of the
symmetry
\be
{\rm pr}^{(1)} K = K + \Phi^x \frac{\partial}{\partial v_x}
\ee 
with
\be
\Phi^x = \frac{d\Phi}{dx} - v_x \frac{d\xi}{dx} 
= \frac{d}{dx} \left( \Phi -  v_x \xi \right) + \xi v_{xx}.
\ee
Since both $v(x)$ and $\tilde{v}(\tilde{x})$ are solutions,
$\Delta(\tilde{v},\tilde{x},\tilde{v}_x)=0$ and 
$\Delta(v,x,v_x)=0$ for arbitrary $\varepsilon$
and we arrive at
\be
\left. {\rm pr}^{(1)} K  (\Delta) \right|_{\Delta=0} \,=\,0.
\label{PRO_ZERO}
\ee


\vs

The requirement of transforming only among solutions 
leads to a number of {\em linear} partial
differential equations for the unknown functions $\Phi$ and $\xi$
(which may contain, in general, several components)
by equating the coefficients of each monomials in $v$, $v_x$ etc.
with zero.
Eq.(\ref{PRO_ZERO}) is a property of the transformation
and therefore is fulfilled for any solution of the original
equation. The $\Delta=0$ constraint is important, it reduces
the dimensionality of the space of the coefficient functions,
usually by expressing one of the highest oder partial
derivatives in terms of others.
The resulting system of equations  in principle can be solved (being linear). 
The corresponding
vector field $K$ will then contain in general a number of constants
or undetermined functions selecting out classes of transformations
which generate new solutions from a given particular solution
(\ref{PD}). The method of prolongation ensures us that we
have considered all possible transformations.

\vs \vs
\noindent
{\bf \sc New solutions}
\vs

Now we rewrite equation (\ref{ANAL_FLOW}) with notations of
the method of prolongation:
\ba
\Delta &=& v_{\LO} + v v_r = 0, \NL
K &=& \Phi(v,\LO,r) \frac{\partial}{\partial v}
+ T(v,\LO,r) \frac{\partial}{\partial \LO}
+ R(v,\LO,r) \frac{\partial}{\partial r}, \NL
{\rm pr}^{(1)} K &=& K + \Phi^{\LO} \frac{\partial}{\partial v_{\LO}}
+ \Phi^r \frac{\partial}{\partial v_r}.
\ea

Applying the prolonged Killing vector field to the nonlinear equation
to be solved we get
\be
{\rm pr}^{(1)} K \Delta = \Phi^{\LO} + v \Phi^r + v_r \Phi = 0.
\label{PRO}
\ee
The prolonged coefficients of the Killing vector field are
\ba
\Phi^{\LO} &=& (\Phi_{\LO} + v_{\LO} \Phi_v) -
(T_{\LO}v_{\LO} + T_v v_{\LO}^2 ) - (R_{\LO} v_r + R_v v_{\LO}u_r), \NL
\Phi^r &=& (\Phi_r + v_r \Phi_v) -
(T_r v_{\LO} + T_v v_{\LO}u_r ) - (R_r v_r + R_v v_r^2 ).
\ea
Replacing it to the prolongation condition eq.(\ref{PRO}), we arrive at
\be
{\rm pr}^{(1)}K \left. (\Delta) \right|_{\Delta=0} \, = \,
\Phi_{\LO} + v \Phi_r + v_r( \Phi - R_{\LO} )
+ v v_r (T_{\LO} - R_r) + v^2 v_r T_r = 0.
\ee
This has to be fulfilled for arbitrary $v(\LO,r)$.
In this equation, since we have already used the $\Delta = 0$
constraint (the original equation to be solved) for eliminating
$v_{\LO} = - v v_r$, each coefficient in the polynomial of $v$
and $v_r$ vanishes
separately, leading to a number of {\em linear} partial differential
equations:
\be
\Phi_{\LO} = 0, \qquad
\Phi_r     = 0, \qquad
\Phi       = R_{\LO}, \qquad
T_{\LO}    = R_r, \qquad
T_r        = 0.
\ee
To these equations (being linear) a general solution can be given:
\ba
T(v,\LO,r)    &=& b(v) + a(v) \LO, \NL
R(v,\LO,r)    &=& c(v) + a(v) r + \phi(v) \LO, \NL
\Phi(v,\LO,r) &=& \phi(v).
\ea
There are four undetermined functions in this general solution to
the Killing vector problem, hence there are four general classes
of transformations leaving the fulfillment of the original equation
untouched. The corresponding vector fileds, which generate infinitesimal
transformations, are
\ba
K_1 &=& \phi(v) \left( \frac{\partial}{\partial v} + \LO 
\frac{\partial}{\partial r} \right), \NL
K_2 &=& a(v) \left( r \frac{\partial}{\partial r} + 
\LO \frac{\partial}{\partial \LO} \right), \NL
K_3 &=& b(v) \frac{\partial}{\partial \LO}, \NL
K_4 &=& c(v) \frac{\partial}{\partial r}.
\ea
A general finite transformation of any solution of $\Delta = 0$ 
is given by
\be
(\tilde{v}, \tilde{\LO}, \tilde{r}) = 
{\rm exp} \left( \epsilon_1 K_1 + \epsilon_2 K_2 + \epsilon_3 K_3
+ \epsilon_4 K_4 \right) (v,\LO,r).
\ee
Three of these transformations are quite trivial:
$K_4$ generates translation of the variable $r$ by $c(v)$,
$K_3$ generates translation of the variable $\LO$ by $b(v)$,
and $K_2$ a strecth of both $r$ and $\LO$ by $e^{a(v)}.$
Only $K_1$ involves a transformation of $v$ as well:
\ba
\tilde{v} &=& v + \psi(v), \NL
\tilde{\LO} &=& \LO, \NL
\tilde{r} &=& r + \LO \psi(v),
\ea
with
\be
\psi(v) = \int \frac{dv}{\Phi(v)}.
\ee
Combination of these transformations generates from the simple
particular solution $v = r/\LO$ the following (implicit)
solution
\be
v = \frac{ r + A(v)}{\LO + B(v)}.
\label{IMPL}
\ee
Fitting this form to an initial transverse velocity profile,
leads to
\be
v_0 = v(0,r) = \frac{r + A(v_0(r))}{B(v_0(r))},
\ee
which is equivalent to
\be
r = v_0 B(v_0) - A(v_0) = f(v_0).
\ee
Finally we arrive at the general solution
\be
r - v \LO = f(v),
\ee
with $f(v)$ being the inverse function of the initial
transverse velocity profile $v_0(r)$. 
The general solution is implicitely given by
\be
v = v_0(r - v \LO).
\label{IMP_SOL}
\ee
It is easy to verify that this implicit form still satisfies the
original equation (\ref{ANAL_FLOW}).

\vs
Finally let us consider some possible initial transverse velocity
profiles. For the linear case,
\be
v_0(r) = \alpha r,
\ee
we obtain
\be
v = \alpha (r - v \LO) 
\ee
which can be resolved to the explicit expression
\be
v = \frac{ \alpha r}{1 + \alpha \LO} = \frac{r}{\LO + 1/\alpha}.
\ee
presented already in \cite{Biro}.
Initial profiles with nonlinear or non-invertable functions
can be resolved only numerically, but a solution of implicit
equations are usually faster and more stable than algorithms for
solving partial differential equations numerically.

\vs
{\bf \sc Flow at SPS}
\vs

A particularly interesting enterprise is to start with a
transverse flow profile conjectured from experimental data
and evolve the flow pattern backwards in (longitudinal) time.
For the CERN SPS heavy ion experiments Ster, Cs\"org\H{o} and L\"orstad
have proposed recently such a profile based on one-particle
transverse momentum spectra and HBT correlation measurments
\cite{Ster}. The reconstructed flow in the notations of the
present paper can be written as,
\be
\gamma v = \sh \eta_t = \langle u_t \rangle \frac{r_t}{R_G} = \alpha r,
\label{REC_FLOW}
\ee
at a late time instant $\LO_0=6$ fm/c. The fit by Ster et. al. 
concludes at $\langle u_t \rangle = 0.55 \pm 0.06 $
and $R_G = 7.1 \pm 0.2$ fm. This gives rise to the following proportionality
coefficient: $\alpha  = 0.0775 \pm 0.01$.
This suggestion has improved on the linear transverse flow
assumption, it nowhere exceeds the speed of light.

\vs
Fig.1 shows the reconstructed relativistic transverse velocity
profiles ending at the one suggested by Ster et.al.
It was assumed that in the final stage the phase transition
just ended in an overall hadronic phase.

\vs
As insertions show the mixed phase should have started then
as early as $\LO = 1.14$ fm/c, but the quark matter part
was decreased to $1/3$ on the average soon ($\LO = 2.4$ fm/c).
The velocity profiles do not differ from linear qualitatively
in the range of interest. It is more pronounced how the
radius of the mixed phase expands from $R_0 = 5.4$ fm to
the reconstructed final stage with $R = 7.1$ fm.


\vs

\begin{figure}
\vskip -0.4in
\epsfxsize=3.0in
\epsfysize=4.0in
\centerline{\rotate[r]{\epsffile{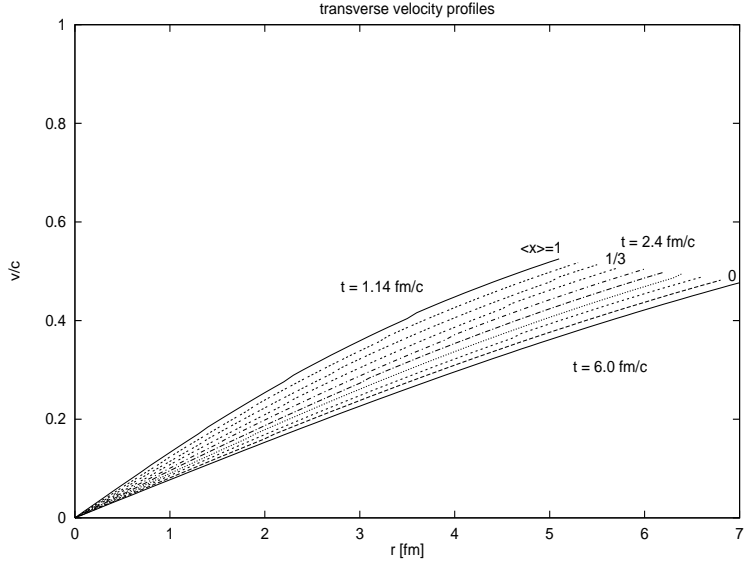}}}
\vskip -0.05in
\caption[]{
 \label{figure1}
Relativistic transverse velocity profiles from the
prolonged analytic solution. The curves are results of a 
time backward calculation from a conjectured final state at
$\LO = 6$ fm/c with a radius of $R = 7.1$ fm. The average
portion of the quark matter in the mixed phase is indicated
at the respective values of $0$, $1/3$ and $1$. 
}
\end{figure}

\vs
{\bf \sc Conclusion}
\vs

With the help of the method of prolongation we explored the
symmetries of Euler's equation for a relativistic transverse
flow at the softest point (zero comoving gradient for pressure).
We arrived at an implicit solution (\ref{IMP_SOL}) which
presents a correspondence between the initial transverse
velocity profile and that at arbitrary longitudinal proper
time. Although in a general case an implicit equation
can only be solved numerically, it is a more elegant, stable
and concise way to solve the flow problem than the numerical
integration of partial differential equations.

With this perspective a transverse flow pattern obtained
from measurements in heavy ion experiments at CERN SPS,
where probably the mixed quark-gluon and hadronic phase
has been realized, can also be re-calculated at the
beginning of phase conversion (pure quark matter).
In turn this might help for obtaining an improved insight
into further characteristics (temperature, effective masses)
of the quark matter at CERN.

\vspace{1.0cm}
\noindent
{\Large {\bf  Acknowledgements}}

This work is part of a collaboration between the
Deutsche Forschungsgemeinschaft and the Hungarian Academy of
Science (project No. DFG-MTA 101/1998) and has been
supported by the Hungarian National Fund for Scientific
Research OTKA (project No. T019700 and T029158),
as well as by the Joint Hungarian-American Technology Fund
TeT-MAKA (no. JF.649).

\vspace{0.5cm}


\end{document}